\documentclass[aps,prl,twocolumn,superscriptaddress,floatfix,nofootinbib,showpacs,amsmath,amssymb,altaffilletter]{revtex4-2}
\pdfoutput=1

\UseRawInputEncoding

\usepackage{amsmath}
\usepackage{amsfonts}
\usepackage{amssymb}
\usepackage{graphicx}
\usepackage{bm}
\usepackage{subfigure}
\usepackage{url}
\usepackage[hyperindex]{hyperref}
\usepackage{color}
\usepackage[ddmmyy,24hr]{datetime}
\usepackage{bigdelim}
\usepackage{booktabs}
\usepackage{dcolumn}
\usepackage{multirow}
\usepackage{subfigure}
\usepackage{physics}
\usepackage{cancel}
\usepackage{stackrel}
\usepackage{paralist}
\usepackage{xspace}
\usepackage{slashed}
\usepackage{cancel}
\usepackage{todonotes}
\usepackage{enumerate}
\usepackage{float}
\usepackage{fullpage}
\usepackage{ulem}
\usepackage{wasysym}
\usepackage{comment}
\usepackage{bbold}
\usepackage{orcidlink}

\usepackage{feynmp}
\newcommand{\nua}[1]{\ensuremath{\rlap{\kern-2.5pt\ensuremath{\overset{\scriptscriptstyle(-)}{\phantom{\nu}}}}{\ensuremath{{\nu}_{#1}}}}}

\usepackage{enumitem}

\begin{document}

\title{Possible solution to the gallium anomaly moving beyond the leptonic wave function factorization}

\author{M. Cadeddu \orcidlink{0000-0002-3974-1995}}
\email{matteo.cadeddu@ca.infn.it}
\affiliation{Istituto Nazionale di Fisica Nucleare (INFN), Sezione di Cagliari,
	Complesso Universitario di Monserrato - S.P. per Sestu Km 0.700,
	09042 Monserrato (Cagliari), Italy}

\author{N. Cargioli \orcidlink{0000-0002-6515-5850}}
\email{nicola.cargioli@ca.infn.it}
\affiliation{Istituto Nazionale di Fisica Nucleare (INFN), Sezione di Cagliari,
	Complesso Universitario di Monserrato - S.P. per Sestu Km 0.700,
	09042 Monserrato (Cagliari), Italy}

\author{F. Dordei \orcidlink{0000-0002-2571-5067}}
\email{francesca.dordei@cern.ch}
\affiliation{Istituto Nazionale di Fisica Nucleare (INFN), Sezione di Cagliari,
	Complesso Universitario di Monserrato - S.P. per Sestu Km 0.700,
	09042 Monserrato (Cagliari), Italy}

\author{L. Ferro \orcidlink{0009-0002-1698-3710}}
\email{luca.ferro@ca.infn.it}
\affiliation{Dipartimento di Fisica, Universit\`{a} degli Studi di Cagliari,
	Complesso Universitario di Monserrato - S.P. per Sestu Km 0.700,
	09042 Monserrato (Cagliari), Italy}
\affiliation{Istituto Nazionale di Fisica Nucleare (INFN), Sezione di Cagliari,
	Complesso Universitario di Monserrato - S.P. per Sestu Km 0.700,
	09042 Monserrato (Cagliari), Italy}

\author{C. Giunti \orcidlink{0000-0003-2281-4788}}
\email{carlo.giunti@to.infn.it}
\affiliation{Istituto Nazionale di Fisica Nucleare (INFN), Sezione di Torino, Via P. Giuria 1, I--10125 Torino, Italy}

\author{M. Pitzalis \orcidlink{0009-0003-1854-0339}}
\email{matilde.pitzalis@ca.infn.it}
\affiliation{Dipartimento di Fisica, Universit\`{a} degli Studi di Cagliari,
	Complesso Universitario di Monserrato - S.P. per Sestu Km 0.700,
	09042 Monserrato (Cagliari), Italy}
\affiliation{Istituto Nazionale di Fisica Nucleare (INFN), Sezione di Cagliari,
	Complesso Universitario di Monserrato - S.P. per Sestu Km 0.700,
	09042 Monserrato (Cagliari), Italy}

\begin{abstract}

For over thirty years, a $\sim20\%$ deficit, now exceeding $5\sigma$, has persisted between measured and predicted neutrino capture rates on $^{71}$Ga, as observed in radioactive source experiments (namely GALLEX, SAGE, and more recently BEST) using $^{51}$Cr and $^{37}$Ar. This long-standing discrepancy, referred to as the gallium anomaly, has posed a significant challenge to our understanding of both experimental methods and theoretical predictions. In this work, we revisit the theoretical calculation of the neutrino capture cross-section by moving beyond the standard treatment of the leptonic wave functions, revealing limitations in the commonly used factorization approach based on the detailed balance principle. Incorporating phenomenologically constrained Gamow-Teller transition densities, able to correctly reproduce the precisely measured half-life of $^{71}{\textrm{Ge}}$, we find that the revised cross-section can be significantly reduced,
potentially resolving the gallium anomaly without invoking new physics.
\end{abstract}

\maketitle

\textit{\textbf{Introduction.}}---
Among the open questions in the neutrino sector, a particularly persistent and peculiar anomaly is associated with the response of $^{71}\mathrm{Ga}$ to electron neutrinos, commonly referred to as the \textit{gallium anomaly}~\cite{Elliott:2023cvh, Giunti:2010zu}. The anomaly originates from the charge-current (CC) interaction \mbox{$\nu_e + ^{71}\mathrm{Ga} \rightarrow ^{71}\mathrm{Ge} + e^-$}, a reaction historically exploited for solar neutrino detection. It was originally observed in the GALLEX~\cite{ANSELMANN1992376,1999127,Kaether:2010ag,GNO:2005bds} and SAGE~\cite{SAGE:2009eeu,doi:10.1142/9789811204296_0002} calibration campaigns, which reported measured neutrino capture rates on $^{71}$Ga significantly lower than theoretical expectations when using $^{51}$Cr and $^{37}$Ar artificial neutrino sources. More recently, the BEST experiment~\cite{Barinov:2021asz,PhysRevC.105.065502} has further confirmed and strengthened the anomaly at the $\sim5\sigma$ level~\cite{Cadeddu:2025pue}.
Taken together with other short-baseline neutrino anomalies~\cite{Mention:2011rk,MicroBooNE:2022sdp,Giunti:2022btk,Berryman:2019hme}, this result has prompted renewed discussions of possible new physics scenarios beyond the Standard Model~\cite{Giunti:2006bj,Giunti:2019aiy,Gariazzo:2015rra,Gonzalez-Garcia:2015qrr,Diaz:2019fwt, Boser:2019rta,Dasgupta:2021ies,Giunti:2023kyo,Abazajian:2012ys,Acero:2022wqg,Farzan:2023fqa}. 
However, the simplest and most considered explanation, consisting of short-baseline active-sterile neutrino oscillation, is in tension with the results of reactor antineutrino experiments
\cite{Giunti:2021kab,Giunti:2022btk,STEREO:2022nzk,Machikhiliyan:2022muh,PROSPECT:2024gps}, with the solar neutrino bound
\cite{Goldhagen:2021kxe,Gonzalez-Garcia:2024hmf}, the recent results of MicroBooNE using accelerator neutrino beams~\cite{Microboone} and the KATRIN neutrino mass measurement
\cite{KATRIN:2025lph}. The gallium anomaly, especially, continues to stand out as a stubborn inconsistency that seems to call for explanations other than sterile-neutrino scenarios~\cite{Huber:2025nat}.
Thus, a careful reassessment of the assumptions underlying the calculation of the theoretical cross section for the inverse beta decay (IBD) reaction \mbox{$\nu_e + ^{71}\mathrm{Ga} \rightarrow ^{71}\mathrm{Ge} + e^-$} is in order. 

The theoretical prediction of the cross section, initiated in the foundational work of Bahcall~\cite{Bahcall:1997eg}, has undergone several refinements over the years~\cite{Elliott:2023xkb,Haxton:2025hye,Kostensalo:2019vmv,Brdar:2023cms,Giunti:2022xat,Huber:2022osv,Cadeddu:2025pue}. Despite these efforts, a full resolution of the anomaly has remained elusive.
Building upon the framework developed in Ref.~\cite{Cadeddu:2025pue}, here we extend the theoretical treatment by critically reassessing the use of the \textit{detailed-balance} (db) principle~\cite{Williams,Blatt:1952ije, Alvarez:1949zz}. Under microscopic reversibility (time-reversal invariance), the latter relates a process and its inverse once phase-space and spin-degeneracy factors are properly included. 
This symmetry has historically allowed hard-to-predict rates to be inferred from their inverse reactions. 
However, in high-precision applications, uncritical reliance on the detailed balance could introduce biases.
In the context of IBD, the detailed-balance approximation, which assumes a strict correspondence between neutrino-induced reactions and electron capture rates~\cite{ Bahcall:book}, has been so far used assuming a factorization of nuclear and leptonic matrix elements. Our work shows that this factorization may introduce significant bias in the theoretical prediction, even when accurate bound and continuum lepton wave functions and nuclear dynamics are accounted for.
Our alternative approach aims to provide a more accurate and self-consistent theoretical framework for interpreting gallium-based and, more broadly, neutrino CC interaction experiments, potentially solving the gallium anomaly without invoking new physics.\\

\textit{\textbf{Detailed-balance approximation.}}---The ground-state IBD cross section for the \mbox{$\nu_e + ^{71}\textrm{Ga} \to  ^{71}{\textrm{Ge}} + e^-$} reaction is\footnote{Natural units with $\hbar = c = 1$ are adopted throughout.}~\cite{BahcallRevModPhys, Bahcall:1997eg,Bahcall:book,Bahcall1964}
\begin{equation}
\sigma_\text{gs}=\frac{G_F^2\,|V_{ud}|^{2}\,g^2_{A}\,}{\pi (2J_\text{Ga}+1)}\sum_{j}{ {p^j_e E^j_e}\,
\left|\mathcal{H}_j^{\rm IBD}\right|^2 \,\mathcal{B}(E^j_e)},
\label{eq:sigma_IBD}
\end{equation}
where $G_F$ denotes the Fermi constant, $V_{ud}=0.9743$ is the CKM matrix element~\cite{ParticleDataGroup:2024cfk}, and $g_A=1.2764$ is the axial-vector coupling~\cite{PERKEO3}. Here, the emitted electron's momentum and energy are indicated by $p_e$ and $E_e$, respectively, $J_\text{Ga}=3/2$ is the nuclear spin of the ground state of $^{71}$Ga and $\mathcal{H}^{\rm IBD}$ is the IBD process transition amplitude. The sum spans all possible electron energies $E_e^j$, each weighted by the branching ratio $\mathcal{B}(E^j_e)$ of the neutrino source.

Following the detailed-balance principle, as discussed in Ref.~\cite{Cadeddu:2025pue}, $\sigma_\text{gs}$ can be determined exploiting the nearly-exact inverse process, namely the $e^-_{\text{bound}} + {}^{71}\text{Ge} \rightarrow {}^{71}\text{Ga} + \nu_e$ electron capture (EC), taking advantage of the well-determined half-life $t_{1/2}$ of $^{71}{\textrm{Ge}}$~\cite{PhysRevC.31.666,Collar:2023yew,PhysRevC.109.055501,newGelifetime}.
Under this assumption, the IBD cross section is given by~\cite{BahcallRevModPhys, Bahcall:1997eg,Bahcall:book,Bahcall1964,Elliott:2023xkb}
\begin{equation}
\sigma^{\rm db}_{\mathrm{gs}} \!=\! \!\frac{2\pi^2\! \ln2}{f_\text{EC}\,t_{1/2}}\!\left(\frac{2J_\text{Ge}\!+\!1}{2J_\text{Ga}\!+\!1} \right) \!
\!\sum_{j}p^j_e E^j_e
\mathcal{F}(E_e,Z,r_0)\mathcal{B}(E^j_e),
\label{eq:sigmaIBDInt}
\end{equation}
where $J_\text{Ge}=1/2$ is the nuclear spin of the ground state of $^{71}$Ge, while $f_\text{EC}$ is the phase space factor for an allowed electron capture and depends mainly on the $1s$ shell bound electron wave function~\cite{BahcallRevModPhys, Bahcall:book}. Finally, \mbox{$\mathcal{F}(E_e,Z,r_0)$} is the Fermi function~\cite{Fermi:1933jpa}, which corrects for the distortion of the outgoing electron wave function mainly due to the Coulomb potential of the daughter nucleus, with $Z$ being its atomic number while $r_0$ is often selected as the origin, $r_0 = 0$.

Invoking the detailed-balance principle is useful to avoid the complicated calculation of $\mathcal{H}^{\rm IBD}$, which requires knowledge of both the leptonic wave functions, $\psi$, and the nuclear ones, $\Psi$, namely
\begin{align}
    \mathcal{H}_j^{\rm IBD}&\!=\!\int \psi^{j\,*}_{e}(\mathbf{r})\,
\Psi^*_{{}^{71}\text{Ge}}(\mathbf{r})\,\hat{H}_{\text{GT}}\,\psi^{j}_{\nu}(\mathbf{r})\,
\Psi_{{}^{71}\text{Ga}}(\mathbf{r})\,\text{d}\mathbf{r}\,,
\label{eq:Htotibd}
\end{align}
with $\hat{H}_{\text{GT}}$ denoting the Gamow-Teller Hamiltonian.
In order to do so, the approximation that the leptonic wave functions, $\psi_{e,\nu}(\mathbf{r})$, can be factorized out is necessary, invoking the fact that they are expected to be almost constant within the nuclear environment.
This allows one to isolate the IBD nuclear matrix element, $\mathcal{M^{\rm IBD}_{\rm nuc}}$, after factorizing the lepton and nuclear parts
\begin{align}
    \mathcal{H}_j^{\rm IBD}&\simeq\psi^{j\,*}_{e}(r_0)\psi_{\nu}^{j}(r_0)\int
\Psi^*_{{}^{71}\text{Ge}}(\mathbf{r})\,\hat{H}_{\text{GT}}\,
\Psi_{{}^{71}\text{Ga}}(\mathbf{r})\,\text{d}\mathbf{r}\nonumber\\
    &=\psi^{j\,*}_{e}(r_0)\psi_{\nu}^{j}(r_0)\,\mathcal{M^{\rm IBD}_{\rm nuc}}\label{eq:Htotfactor}\,.
\end{align}
Exploiting the detailed-balance principle, the IBD nuclear matrix element can be equated with that of the inverse reaction, namely
\begin{equation}
|\mathcal{M}^{\rm IBD}_{\rm nuc}|^2\stackrel{\rm db}{=}|\mathcal{M}^{\rm EC}_{\rm nuc}|^2\,. 
\label{eq:dbprinciple}
\end{equation}
The advantage of this procedure lies in the fact that the latter can be easily extracted from the averaged measurement of the germanium half-life $t_{1/2}^{\rm exp} = 11.465 \pm 0.003~\text{d}$~\cite{PhysRevC.31.666,Collar:2023yew,PhysRevC.109.055501,newGelifetime,Cadeddu:2025pue}.\\

In Ref.~\cite{Cadeddu:2025pue}, we performed a comprehensive evaluation of the IBD ground-state cross section under the detailed-balance framework using this factorization procedure, implementing, for the first time, nuclear-density averaged electron wave functions for both the IBD and EC processes. For these calculations, we developed a dedicated numerical tool based on the \texttt{RADIAL} package~\cite{radial}, which has been cross-checked with other available tools like \textit{e.g.} \texttt{GRASP}~\cite{Grasp2018}, enabling precise evaluations of both bound and continuum electron wave functions for finite-size nuclei.
While doing so, we incorporated the most recent inputs, including the measurement of the $^{71}\text{Ge}$ nuclear charge radius~\cite{ExpNuclearRadius}, updated atomic-shell capture probabilities~\cite{Bambynek,Collar:2023yew}, and the revised $Q$-value of the transition~\cite{Qvalue}. Despite these improvements,  a significant anomaly remains; namely, $4.7\sigma\,(5.5\sigma)$ depending on the excited-state contribution that is considered~\cite{KrofcheckPhysRevLett.55.1051,FREKERS2011134}.\\


\textit{\textbf{Beyond the factorization scheme.}}---
In this work, we stress that separating the leptonic wave functions from the nuclear part is rigorously justified only in the limit where the leptonic wave functions can be treated as spatially constant over the nuclear volume. Once this approximation is relaxed, the amplitude no longer factorizes into a purely nuclear matrix element times a leptonic factor: the leptonic radial functions remain inside the nuclear-volume integral, so an isolated $\mathcal{M}^{\rm IBD}_{\rm nuc}$ cannot be defined as in Eq.~(\ref{eq:Htotfactor}). As a consequence, the detailed-balance relation invoked in Eq.~(\ref{eq:dbprinciple}) cannot be applied in its standard form to map the IBD nuclear matrix element onto that extracted from EC. Moreover, electron capture is not the exact time-reversed process of neutrino capture, most notably because it involves a bound initial-state electron rather than an outgoing continuum electron, so 
$\mathcal{H}^{\rm IBD}$ cannot be simply inferred from the EC decay rate.

We can overcome this limitation by having in hand the exact radial solutions of the Dirac-Hartree-Fock-Slater (DHFS) equation for the electron wave function and avoiding factorization. 
In this sense, one should reconsider the full matrix element in Eq.~(\ref{eq:Htotibd}), and notice that a more accurate treatment should account for the dependence of the full IBD transition amplitude on the so-called weak transition density, $\rho_{\rm TD}(\textbf{r})$, defined as
\begin{equation}
    \rho_{\rm TD}(\mathbf{r})=\Psi^*_{^{71}\text{Ge}}\,(\mathbf{r})\hat{H}_{\text{GT}}\,\Psi_{^{71}\text{Ga}}\,(\mathbf{r})\, .
    \label{rhoTD}
\end{equation}
The latter describes the part of the nuclear wave functions involved in the transition process, effectively converting a neutron into a proton.
This allows for a more rigorous integration of the leptonic contribution with the nuclear matrix element, without the need to separate the two.\\

Without invoking the db principle, the ground-state IBD cross section reads as in Eq.~(\ref{eq:sigma_IBD}).
Writing out the Gamow-Teller Hamiltonian $\hat{H}_{\rm GT}$, for the specific $^{71}\textrm{Ga}\rightarrow\,^{71}\textrm{Ge}$ process, as well as the leptonic wave functions,
we can recalculate the full transition amplitude in Eq.~(\ref{eq:Htotibd}) readapting the formalism developed for the $\beta$-decay process in Refs.~\cite{konopinski1966theory,KOSHIGIRI1979301,Raman:1978qta,10.1093/ptep/ptab069}. Retaining only the dominant axial-spatial component relevant for an allowed transition, one gets
\begin{align}
|\mathcal{H}^{\rm IBD}|^2&= \left(4 \pi\right)^2 \left(|I_1|^2 + |I_2|^2\right)\,,
\label{eq:transition}
\end{align}
where
\begin{align}
I_1 &= \int dr\,r^2 \rho_{\rm TD}(r)\big[g_{-1}(r) j_0(q\,r) +\tfrac13 f_{-1}(r) j_1(q\,r)\big],\end{align}
and
\begin{align}
I_2 &= \int dr\,r^2 \rho_{\rm TD}(r)\big[ f_{1}(r)j_0(q\,r)-\tfrac13 g_{1}(r)j_1(q\,r)\big],
\label{eq:transitionI}
\end{align}
with $j_0$ and $j_1$ being the spherical Bessel functions of order $0$ and $1$, respectively, describing the neutrino wave functions and $q$ the neutrino momentum. Furthermore, $g_{\kappa}(r)$ and $f_{\kappa}(r)$ are the large and small Dirac electron wave-function radial components, respectively, for a given eigenvalue $\kappa$, which assumes values $\kappa = \pm 1$ for allowed decays, as defined in Ref.~\cite{10.1093/ptep/ptab069}.

\begin{table*}[t]
\resizebox{0.95\textwidth}{!}
{\renewcommand{\arraystretch}{1.8} 
\begin{tabular}{l|c|c|c|c|c|c|c}
Model & Best fit parameter values $\Theta$ & $\sigma_{\rm gs,^{51}\text{Cr}}(\Theta)\, [10^{-45}\, \text{cm}^2]$ & $\sigma_{\rm gs,^{37}\text{Ar}}(\Theta)\, [10^{-45}\, \text{cm}^2]$ & \hspace{0.1cm}$t_{1/2}(\Theta)\, [\text{d}]$\hspace{0.1cm} & \hspace{0.1cm}$\chi^2_{\rm IBD}$\hspace{0.1cm} & $\chi^2_{\rm EC}$ & GA solved\\
\hline
\hline
SG & \{0.00094,-0.005,3.42\}& 5.27 & 6.30 & 11.515 &  8.68 & 0.24 & no \\
$\rm DG$ &  \{-0.0231,-0.0802,3.5119,1.855,2.16,2.16\} & 4.42 & 5.13 & 11.462  &0.014 &0.0005 & yes \\
$\rm mDG$ &  \{-0.1498,0.3462,1.192,1.3793,2.6192,0.466405\} & 4.50 & 5.24 & 11.465  & 0.037 & 0.00003 & yes \\[+0.15ex]
\multirow{2}{*}{$\rm mTG$} &  \{3.514009,4.53567,-1.5007,1.9457,1.3809, & \multirow{2}{*}{4.39} & \multirow{2}{*}{5.10} & \multirow{2}{*}{11.463}  & \multirow{2}{*}{0.044} & \multirow{2}{*}{0.0002} & \multirow{2}{*}{yes} \\[-1.5ex]
 &1.14085,0.0029,0.8595,1.513\} &  &  &   &  & & 
\end{tabular}
}

\caption{Numerical values of the theoretical IBD ground state cross section $\sigma_{\rm gs,^{51}\text{Cr}}(\Theta)$ and $\sigma_{\rm gs,^{37}\text{Ar}}(\Theta)$ for the $^{51}$Cr and $^{37}$Ar sources, respectively, and of the half-life of $^{71}$Ge $t_{1/2} (\Theta)$ obtained with the best fit parameter values in the second column for different transition density parametrizations. For completeness, we report the corresponding $\chi^2$'s, as defined in Eqs.~(\ref{chi2neu}) and (\ref{chi2EC}). For comparison, the experimental IBD ground state cross section and half-life of germanium are 
$\sigma^{\rm exp}_{\rm gs,^{51}\text{Cr}}\, [10^{-45}\, \text{cm}^2]=4.44(33)$, $\sigma^{\rm exp}_{\rm gs,^{37}\text{Ar}}\, [10^{-45}\, \text{cm}^2]=5.21(72)$ and $t_{1/2}^{\rm exp}\, [\text{d}]=11.465(3)$. In the last column we indicate if the corresponding parametrization is able to solve the gallium anomaly (GA).}
\label{tab:summaryTable}
\end{table*}

This procedure significantly improves the rigor of the neutrino-induced cross-section calculation. However, it requires a precise knowledge of the transition density from nuclear theory.
Unfortunately, a first-principles calculation with controlled uncertainties remains challenging~\cite{King:2020pza}.

The same formalism can also be applied to the electron-capture rate, which is related to the germanium half-life as~\cite{Bambynek, Behrens,BahcallEC1962,Bahcall1964,BryskRose}
\begin{align}
\frac{1}{t_{1/2}} =  \frac{G_F^2\,|V_{ud}|^{2}\,g^2_{A}}{\pi \ln 2(2J_\text{Ge}\!+\!1)}
(E^{1s}_{\nu})^2\, 
|\mathcal{H}^{\rm EC}_{1s}|^2 \left[ 1 \!+\! {P_\text{L}+P_\text{M} \over P_\text{K}}\!\right]\,,\label{eq:lambdaEC_iniz}
\end{align}
where $E^{1s}_{\nu}$ is the neutrino energy corresponding to a $1$s shell electron capture and $P_\text{L,M,K}$ are the experimentally measured electron-capture probabilities for the L, M and K shells~\cite{Collar:2023yew, PhysRevLett.125.141301,Bambynek}.\footnote{Here, as well as in the IBD cross section, we account for radiative corrections and the weak magnetism contribution following Ref.~\cite{Elliott:2023xkb}, despite not explicitly written.} Here, $\mathcal{H}^{\rm EC}_{1s}$ is the EC matrix element, which for an allowed electron capture involving $1$s shell electrons, restricted to the leading axial-spatial component, is 
\begin{align}
|\mathcal{H}^\text{EC}_{1\mathrm{s}}|^2&=
\Big|\int dr\, 4\pi\,r^2 \rho_{\rm TD}(r)\big[g_{1\mathrm{s}}(r) j_0(q\,r)\nonumber \\
&+\tfrac13 f_{1\mathrm{s}}(r) j_1(q\,r)\big]\Big|^2\, ,
\label{eq:lambdaEC}
\end{align}
with $g_{1\mathrm{s}}$ and $f_{1\mathrm{s}}$ being the large and small bound electron wave-function radial components. Being the EC transition density the same as that of the IBD $^{71}$Ga($\nu_e$,\,$e^-$)$^{71}$Ge reaction (up to a trivial overall phase), the EC process serves as a crucial validity test. Indeed, since the $^{71}$Ge half-life is experimentally well-constrained, any phenomenological $\rho_{\rm TD}$ capable of resolving the gallium anomaly must simultaneously reproduce the observed germanium half-life.

Here, we assess the impact of the proposed approach using different data-driven phenomenological parametrizations for the transition density. By constraining these models to the precisely measured $^{71}$Ge half-life, we demonstrate their potential to resolve the gallium anomaly.\\


\textit{\textbf{Transition density parametrization.}}---
In general, each transition density parametrization will depend upon a certain set of parameters $\Theta$, in addition to the radial dependence, such that $\rho_{\rm TD}(r,\Theta)$.
We investigate whether (and which) values of $\Theta$ can produce transition densities 
that lead to a theoretical ground-state cross section which solves the gallium anomaly, \textit{i.e.}  
consistent with the values of \( \sigma^{\rm exp}_{\text{gs},\, ^{51}\mathrm{Cr}} = (4.44 \pm 0.33) \times 10^{-45}~\rm{cm}^2 \) and \( \sigma^{\rm exp}_{\text{gs},\, ^{37}\mathrm{Ar}} = (5.21\pm 0.72) \times 10^{-45}~\rm{cm}^2 \), for the $^{51}$Cr and $^{37}$Ar neutrino sources, respectively. These values have been obtained averaging the experimental results reported by the GALLEX, SAGE, and BEST experiments (as listed in Tab. IV of Ref.~\cite{Cadeddu:2025pue}) separately for each neutrino source. 
To isolate the ground-state contribution, we subtract the excited-state component $\Delta_{\text{(p,n)}}$, assigning to it a conservative uncertainty of $\pm5\%$.\footnote{This value is used to include both the $\sim9\%$  contribution obtained from ($^{3}\text{He},^{3}\text{H}$) reaction data~\cite{FREKERS2011134} as well as the possible absence of excited state contributions.} Based on forward-angle (p,n) scattering data~\cite{KrofcheckPhysRevLett.55.1051}, we adopt the values $\Delta_{\text{(p,n)}}=5.3\%$ and 5.8\% for the $^{51}$Cr and $^{37}$Ar sources, respectively.

In this work, we tested different parametrizations, namely a Single Gaussian (SG), a Double Gaussian (DG), and two combinations of polynomial powers of $r$ and Gaussian functions, which we refer to as modified Double Gaussian (mDG) and modified Triple Gaussian (mTG), defined as
\begin{align}
&\rho_{\mathrm{TD}}^{\rm SG}(r,\Theta^{\rm SG}) = A\,e^{-(r-r_a)^2/2a^2} \,,\nonumber\\
&\rho_{\mathrm{TD}}^{\rm DG}(r,\Theta^{\rm DG})\! = \! A\,e^{-(r-r_a)^2/2a^2}\! - \!B\,e^{-(r - r_b)^2/2b^2},\nonumber\\
&\rho_{\mathrm{TD}}^{\rm mDG}(r,\Theta^{\rm mDG})\!=\! Ar\,e^{-(r-r_a)^2/2a^2}\!\!+\!B\, r^2\,e^{-(r - r_b)^2/2b^2}\!,\nonumber\\
&\rho_{\mathrm{TD}}^{\rm mTG}(r,\Theta^{\rm mTG})\! = \! Ar\,e^{-(r-r_a)^2/2a^2}\!\!+\!B\, r^2\,e^{-(r - r_b)^2/2b^2}\!\nonumber\\
&\hspace{2.6 cm}+\!Cr^3e^{-(r - r_c)^2/2c^2}\,,
\label{eq:TD}
\end{align}
where the parameters that are fitted for are \mbox{$\Theta^{\rm SG}\!=\!\{A,r_a,a\}$}, \mbox{$\Theta^{\rm DG}\!=\!\{A,B,a,b,r_a,r_b\}$}, \mbox{$\Theta^{\rm mDG}\!=\!\{A,B,a,b,r_a,r_b\}$} and \mbox{$\Theta^{\rm mTG}\!=\!\{A,B,C,a,b,c,r_a,r_b,r_c\}$}, respectively.

Here, the radii and standard deviations are expressed in $[{\rm fm}]$, while the amplitudes are such that $\rho_{\rm TD}$ is in units of $[{\rm fm}^{-3}]$. The SG and DG forms provide minimal one- and two-lobe descriptions. The mDG and mTG forms were introduced as they mimic, in a minimal analytic way, the typical polynomial-Gaussian structure of products of single-particle wave functions, as for the case of harmonic-oscillator radial wave functions~\cite{PhysRev.117.174}. The powers $r$, $r^2$ and $r^3$ should therefore not be interpreted as a one-to-one assignment to specific shell-model orbitals, but as a compact analytic way of enforcing regular behavior near the origin.

In the $fpg$ valence space~\cite{PhysRevC.80.064323,Kostensalo:2019vmv}, the dominant $p_{3/2}\to p_{1/2}$ component is not the only allowed radial structure by the GT operator. The latter connects proton and neutron single-particle components with the same orbital angular momentum ($\ell$), such as $f_{5/2}\to f_{5/2}$ or $g_{9/2}\to g_{9/2}$. In principle, these additional components present in this space, may interfere coherently with the leading contribution and thereby produce compact nodal profiles, so that the radial transition density is not sign definite.
This possibility is motivated by the
configuration mixing and shell evolution observed in neighboring Ga and Ge
isotopes~\cite{PhysRevLett.104.252502,Kanellakopoulos:2020byf} and by charge-radius systematics indicating shape-polarization effects in
germanium isotopes in the same region~\cite{ExpNuclearRadius}.
We stress that the mDG and mTG forms should not be interpreted as microscopic shell-model transition densities. They are physics-guided phenomenological surrogates designed to test whether compact, radially structured GT densities can make the EC and IBD projections differ. Demonstrating this is the primary goal of the present Letter,  while a microscopic calculation of $\rho_{\rm TD}(r)$ for $^{71}{\rm Ga}\leftrightarrow{}^{71}{\rm Ge}$ is left to future nuclear-structure work.\\
\\
\textit{\textbf{Fit results.}}---
To find the best set of parameters to solve or alleviate the gallium anomaly, given a certain parametrization, we perform a fit using the least-squares function
\begin{align}
\chi_{\rm IBD}^2(\Theta) = \frac{\eta_1^2}{\delta_{\rm \eta_1}^2} + 
\sum_X\left( \frac{\sigma^{\rm exp}_{\text{gs},\, X} - (1+\eta_1) \sigma_{\text{gs},\, X}(\Theta)}{\delta\sigma^{\rm exp}_{\text{gs},\, X}} \right)^2,
\label{chi2neu}
\end{align}
where $X=(^{51}\mathrm{Cr},\,^{37}\mathrm{Ar}$) indicates the neutrino source, $\delta\sigma^{\rm exp}_{\text{gs},\, X}$ is the uncertainty on the corresponding experimental cross section (considering the aforementioned excited state contribution) and $\delta_{\rm \eta_1}=0.003\%$ takes into account the small uncertainty associated to the free electron wave-function calculation as detailed in Ref.~\cite{Cadeddu:2025pue}.

To make the model more realistic, we constrain the Gamow-Teller transition density in Eq.~(\ref{rhoTD}) so that it reproduces the measured $^{71}$Ge electron-capture half-life.
In this case, we define the following \( \chi^2 \)
\begin{align}\nonumber
\chi_{\rm \text{EC}}^2(\Theta) &= \left( \frac{t_{1/2}^{\rm exp}-(1+\eta_2+\eta_3)\cdot t_{1/2}(\Theta)}{\delta{t_{1/2}}} \right)^2 \\
&+ \frac{\eta_2^2}{\delta_{\eta_2}^2} + \frac{\eta_3^2}{\delta_{\rm \eta_3}^2},
\label{chi2EC}
\end{align}
where \( t_{1/2}^{\rm exp} = 11.465 \)\,d and \( t_{1/2}(\Theta) \) is obtained from Eq.~(\ref{eq:lambdaEC_iniz}) using Eq.~(\ref{eq:lambdaEC}). Moreover, $\delta{t_{1/2}^{\rm exp}} = 0.003\,\text{d}$ is the uncertainty on the experimental half-life of \( ^{71}\text{Ge} \), and $\eta_2$ and $\eta_3$ represent nuisance parameters with uncertainties \( \delta_{\rm \eta_2} = 0.007 \) and \( \delta_{\rm \eta_3} = 0.005 \), where the former includes the contribution due to the exchange and overlap corrections (the effect due to the neutrino energy and the electron-capture probabilities is found to be negligible), while the latter takes into account the uncertainty associated to the bound electron wave-function calculation as detailed in Ref.~\cite{Cadeddu:2025pue}.\\

\begin{figure}[t]
    \centering
    \includegraphics[width=\linewidth]{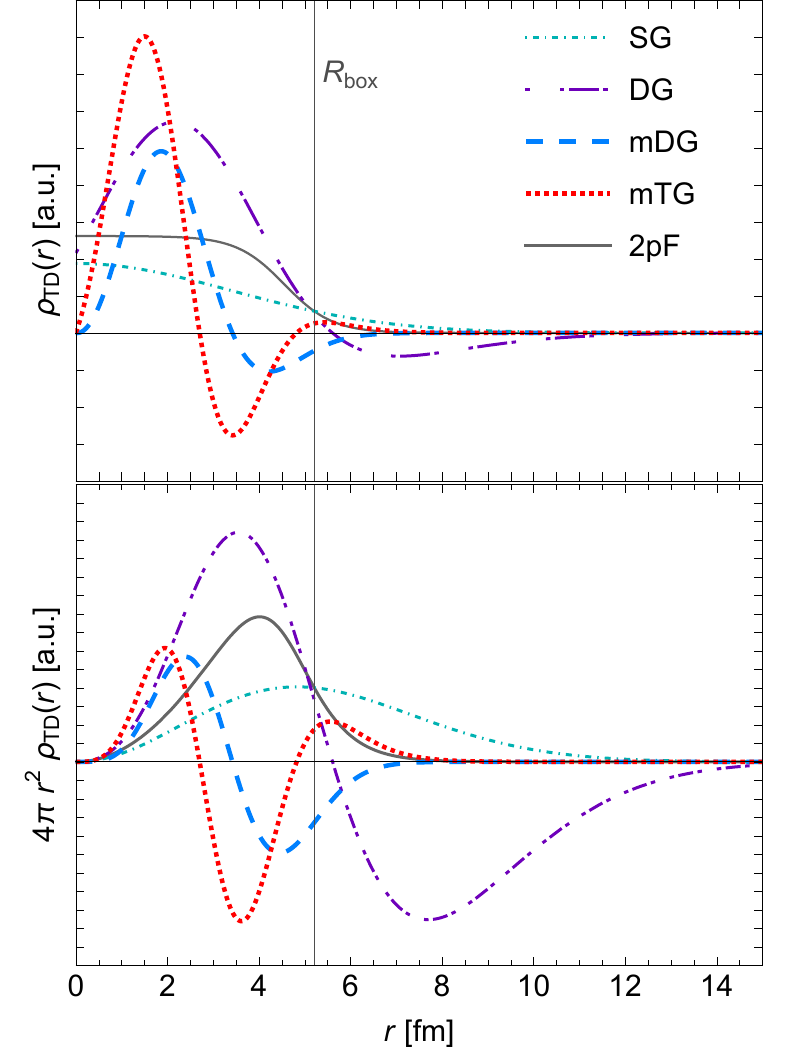}
    \caption{Shape of the transition densities corresponding to the best fit parameters in Tab.~\ref{tab:summaryTable} from the IBD+EC combined fit for the Single Gaussian (SG), the Double Gaussian (DG), the modified Double Gaussian (mDG) and the  modified Triple Gaussian (mTG) parametrization, as well as the two-parameter Fermi (2pF) one. For graphical reasons, the plot is shown in arbitrary units, in order to compare the different results on the same $y$-scale.
    The vertical line indicates the box radius of germanium, $R_{\rm box}=\sqrt{5/3}\,R_{\rm ch}$.}
    \label{fig:plotEC}
\end{figure}

Performing a combined fit, obtained by summing the $\chi^2$ functions in Eqs.~(\ref{chi2neu}) and (\ref{chi2EC}), one may find a transition density parametrization that simultaneously reproduces the experimental half-life of germanium and resolves or alleviates the gallium anomaly by reducing the ground-state cross section up to $\sim20\%$. 
The results of the combined fit are listed in Tab.~\ref{tab:summaryTable} and the corresponding transition densities are shown in Fig.~\ref{fig:plotEC}. 
In the latter, we also report the outcome obtained by adopting a two-parameter Fermi (2pF) shape~\cite{Maximon:1966sqn}
\begin{equation}
\rho_{\mathrm{TD}}^{\rm 2pF}(r,\Theta^{\rm 2pF}) = A \, \dfrac{\rho_0(c,a)}{1+e^{(r-c)/a}}\,,
\end{equation}
\textit{i.e.}, by effectively identifying the weak transition density with the nuclear electric charge distribution. While this ansatz may look reasonable at first sight, it is conceptually unjustified because a charge density is a ground-state electromagnetic observable, whereas $\rho_{\rm TD}$ is a weak transition quantity. 
Notably, even in the case of superallowed Fermi transitions, where the Conserved Vector Current hypothesis relates weak and electromagnetic currents, recent theoretical studies have shown that the charged weak density distinctively deviates from the electric charge distribution, for instance by exhibiting a more pronounced surface-peaking due to nuclear structure effects~\cite{Gorchtein:2023naa, Seng:2023cgl,Seng:2022inj}.
Consequently, relying on the charge distribution to describe the weak transition density is conceptually unjustified and potentially inaccurate, a conclusion that holds even more strongly for Gamow-Teller transitions where no underlying symmetry relates the two distributions. We therefore include it only as a benchmark to quantify the impact of this common simplifying assumption\footnote{The most appropriate baseline for a GT operator is expected to be a surface-peaked transition density rather than the total charge density. We include the 2pF charge distribution only as a conventional reference, corresponding to the simplifying assumption underlying the standard factorized treatment, and not as our best estimate of the physical GT transition density.}. We fix
$\Theta^{\rm 2pF}=\{A=0.588,\;c=4.56\,{\rm fm},\;a=0.52\,{\rm fm}\}$: $c$ and $a$ reproduce the experimental $^{71}$Ge charge radius, $R_{\rm ch}=4.032\pm0.002$~fm~\cite{ExpNuclearRadius}, while $A$ is chosen to match $\mathcal{M}_{\rm nuc}^{\rm EC}$ within the detailed-balance-based approximation. This choice corresponds to the transition-density assumption underlying the latest IBD cross-section prediction in Ref.~\cite{Cadeddu:2025pue}; hence, that prediction is recovered only insofar as $\rho_{\rm TD}(r)$ is taken to follow the given 2pF $^{71}$Ge charge density. In contrast, when confronted with our combined constraints, the resulting ground-state IBD cross section is significantly larger than the experimental one.\\
Remarkably, the DG, mDG and mTG transition densities demonstrate the capacity to fully resolve the gallium anomaly while strictly satisfying the stringent constraint imposed by the precise germanium half-life measurements. This result is intimately connected to the presence of at least one node in the transition density. Thus, as expected, the SG does not solve the tension. 
As clearly visible in Fig.~\ref{fig:plotEC}, the DG parametrization extends significantly beyond the germanium nuclear radius. In contrast, both mDG and mTG resolve the anomaly with compact transition densities localized around the ordinary nuclear surface region, graphically indicated by the 2pF curve and by the vertical line. Therefore, mDG and mTG provide solutions more in line with standard nuclear-structure expectations, whereas the extended DG solution should be regarded only as an illustrative limiting case. Importantly, the compact mDG and mTG examples show that the mechanism does not require anomalously long-ranged transition densities.

A comparable suppression effect has recently been proposed in the context of both $\beta$-decay~\cite{10.1093/ptep/ptab069,Nabi:2019usr} and electron capture processes~\cite{Ravlic:2025wrl}, where the use of refined lepton wave functions combined with particular weak TD's was shown to decrease transition rates by up to 40-50\% in heavy nuclei. These findings lend additional support to the significant cross-section variations reported in the present work.

Although the transition densities adopted here are not unique and rely on specific modeling assumptions, our primary goal is to provide a proof of principle: we demonstrate that physically reasonable transition densities can induce cross-section modifications large enough to remove the gallium anomaly, without violating the tight experimental bounds on the germanium half-life. It is important to emphasize that identifying alternative parametrizations or parameter sets capable of mitigating the anomaly's statistical significance, even if only partially, would already constitute a substantial achievement. 

As a last remark, we note that under the mechanism proposed here, the (p,n) charge-exchange data used to normalize excited-state contributions are referenced to the ground-state matrix element inferred from electron capture. If the present scenario is correct, this normalization would need to be revisited in a dedicated analysis. However, we do not expect this to qualitatively affect our conclusions. A quantitative reassessment is left to future work.\\

\textit{\textbf{Conclusions.}}---
In this work, we critically reassess the validity of the detailed-balance principle in relation to the gallium anomaly, abandoning the usual factorization between leptonic and nuclear currents in the Gamow-Teller transition amplitudes for both the inverse beta decay and the electron capture processes. 
As the detailed balance is no longer applicable, the nuclear part of the interaction can be described by phenomenologically motivated weak transition densities.
By employing exact DHFS electron wave functions alongside data-driven transition densities, we identify different parametrizations that yield a substantial reduction ($\sim$20\%) in the charge-current neutrino capture cross-section on $^{71}$Ga compared to previous estimates, while still preserving consistency with experimental measurements of the electron capture lifetime on $^{71}$Ge. This reduction offers a viable solution to the gallium anomaly, eliminating the need to invoke beyond the Standard Model mechanisms such as sterile neutrinos, which at the moment are very disfavored in light of the recent MicroBooNE and KATRIN results. 

We emphasize that going beyond the standard factorization approximation, by using precise lepton wave functions integrated together with the nuclear transition density, constitutes a necessary but not sufficient condition to account for the observed reduction of the neutrino-capture cross section. Such a reduction is achieved only if the transition density has sufficient radial structure, in particular at least one sign change. Importantly, the compact mDG and mTG examples show that the required sign-changing profile need not be anomalously long-ranged. For these reasons, we strongly encourage the nuclear physics community to intensify efforts toward a reliable and accurate determination of transition densities in these processes, both theoretically and experimentally.\\

\textit{Acknowledgments.}---
We thank G. Co' for useful discussions on shell-model radial wave functions and transition densities.

\bibliography{ref}



\end{document}